\documentclass{PoS}

\newcommand{\barray}{\begin{eqnarray}}
\newcommand{\earray}{\end{eqnarray}}

\newcommand{\beq}{\begin{equation}}
\newcommand{\eeq}{\end{equation}}
\newcommand{\ba}{\begin{array}}
\newcommand{\ea}{\end{array}}
\newcommand{\bea}{\begin{eqnarray}}
\newcommand{\eea}{\end{eqnarray} }
\newcommand{\be}{\begin{eqnarray}}
\newcommand{\ee}{\end{eqnarray} }
\newcommand{\bal}{\begin{align}}
\newcommand{\eal}{\end{align}}
\newcommand{\bi}{\begin{itemize}}
\newcommand{\ei}{\end{itemize}}
\newcommand{\ben}{\begin{enumerate}}
\newcommand{\een}{\end{enumerate}}
\newcommand{\bc}{\begin{center}}
\newcommand{\ec}{\end{center}}
\newcommand{\bt}{\begin{table}}
\newcommand{\et}{\end{table}}
\newcommand{\btb}{\begin{tabular}}
\newcommand{\etb}{\end{tabular}}

\DeclareOldFontCommand{\brianup}{\upshape}{\mathrm}

\DeclareSymbolFont{EUR}{U}{eur}{m}{n}
\SetSymbolFont{EUR}{bold}{U}{eur}{b}{n}
\DeclareSymbolFontAlphabet{\eur}{EUR}

\DeclareSymbolFont{EUB}{U}{eur}{b}{n}
\SetSymbolFont{EUB}{bold}{U}{eur}{b}{n}
\DeclareSymbolFontAlphabet{\eub}{EUB}

\DeclareSymbolFont{AMSb}{U}{msb}{m}{n}
\DeclareSymbolFontAlphabet{\mathbb}{AMSb}

\newcommand{\notyet}[1]{{}}

\DeclareMathSymbol{\varGamma}{\mathord}{letters}{"00}
\DeclareMathSymbol{\varDelta}{\mathord}{letters}{"01}
\DeclareMathSymbol{\varTheta}{\mathord}{letters}{"02}
\DeclareMathSymbol{\varLambda}{\mathord}{letters}{"03}
\DeclareMathSymbol{\varXi}{\mathord}{letters}{"04}
\DeclareMathSymbol{\varPi}{\mathord}{letters}{"05}
\DeclareMathSymbol{\varSigma}{\mathord}{letters}{"06}
\DeclareMathSymbol{\varUpsilon}{\mathord}{letters}{"07}
\DeclareMathSymbol{\varPhi}{\mathord}{letters}{"08}
\DeclareMathSymbol{\varPsi}{\mathord}{letters}{"09}
\DeclareMathSymbol{\varOmega}{\mathord}{letters}{"0A}

\makeatletter\@addtoreset{equation}{section}
\makeatother

\title{Gauge theory of Lorentz group on the lattice}

\ShortTitle{Gauge theory of Lorentz group on the lattice}

\author{\speaker{M.A.Zubkov}\\
        Institute for Theoretical and Experimental Physics\\B.Cherjemushkinskaya - 25, Moscow, Russia\\
        E-mail: \email{zubkov@itep.ru}}

\abstract{The model with the fermions coupled in the non - minimal way to the gauge theory of Lorentz group is considered. The lattice regularization is suggested. It is argued that this model may exist in the phase with broken chiral symmetry and without confinement. We speculate about the possibility that this construction may serve as an origin of the dynamical electroweak symmetry breaking.
}

\FullConference{31st International Symposium on Lattice Field Theory - LATTICE 2013\\
		July 29 - August 3, 2013\\
		Mainz, Germany}

\begin{document}

\section{Introduction}

The scale of the Riemannian gravity (considered as the dynamical theory of metric) is the Plank mass. This is many orders of magnitude larger than the expected scale for the interactions that may  cause the formation of the composite Higgs bosons and provide the dynamical electroweak symmetry breaking. However, one may consider quantum gravity with torsion \cite{Shapiro,Elizalde}. Its dynamical variables are the vierbein (that is related to Riemannian gravity) and the $SO(3,1)$ connection that gives rise to torsion. With the vierbein frozen we come to the gauge theory of Lorentz group \cite{Minkowski,Minkowski_}. This gauge theory is interesting itself. It certainly deserves to be  investigated using nonperturbative lattice methods. Here we present the setup for such investigations.

Moreover, as it will be explained further, this theory may appear to be the constituent of the ultraviolet completion of the Standard Model (SM) responsible for the dynamical electroweak symmetry breaking. Its
scale is not fixed but is assumed to be between $10^3$ TeV and $10^{15}$ GeV while the scale of the dynamical theory of metric is assumed to be given by the Plank mass $\approx 10^{19}$ GeV. Unlike \cite{Z2010_3} we do not introduce the additional technifermions. As in the models of top - quark condensation \cite{topcolor1,Miransky,Simmons} the bilinear combinations of the SM fermions play the role of composite Higgs bosons.

 The dynamical electroweak symmetry breaking can be explained within the effective NJL approximation \cite{NJL} to the ultraviolet completion of the SM. This NJL approximation has the finite ultraviolet cutoff $\Lambda$. In \cite{VZ2013,VZ2012,VZ2013_2,Zubkov:2013zg} it was proposed that the contributions of the microscopic theory due to the trans - $\Lambda$ degrees of freedom cancel the dominant higher loop divergences of the low energy NJL effective theory basing on the analogy with the hydrodynamics \cite{quantum_hydro,hydro_gravity}. Therefore, the one - loop results dominate. Here we follow this proposition. Sure, this is a kind of the fine - tuning. But, is this possible to avoid the fine - tuning while providing the fermion masses from less than $1$ eV for neutrino to about $174$ GeV for the top - quark?

\section{Gauge field}

Let us consider the action of the form $S_g = S_T + S_G$ with
\begin{equation}
S_T = -M_T^2 \int  G d^4x \label{ST}
\end{equation}
and
\begin{eqnarray}
S_G &=&  \beta_1 \int  G^{abcd}G_{abcd}d^4x+\beta_2 \int
G^{abcd}G_{cdab}d^4x +\beta_3 \int G^{ab}G_{ab}d^4x+\beta_4 \int
G^{ab}G_{ba}d^4x\nonumber\\&& +\beta_5 \int G^2d^4x+\beta_6 \int
A^2 d^4x\label{ST2}
\end{eqnarray}
with coupling constants $\beta_{1,2,3,4,5,6}$. Here $G^{ab}_{..\mu\nu} = [D_{\mu},D_{\nu}]$ is the $SO(3,1)$ curvature,  the covariant derivative is $D_{\mu} =
\partial_{\mu} + \frac{1}{4}C_{..\mu}^{ab} \gamma_{[a}\gamma_{b]}$;
$\gamma_{[a}\gamma_{b]} =
\frac{1}{2}(\gamma_{a}\gamma_{b}-\gamma_{b}\gamma_{a})$.
The spin
connection is denoted by $C_{\mu}$.
Indices are lowered and lifted via metric $g$ of Minkowsky space as
usual.
$G^{abcd}=\delta^c_{\mu}\delta^d_{\nu}G^{ab}_{\mu\nu}$, $G^{ac}=G^{abc}_{...b}$, $G =
G^a_a$, $A = \epsilon^{abcd} G_{abcd}$. 
 Actually,
Eq. (\ref{ST2}) is the most general quadratic in curvature action that does
not break Parity \cite{Diakonov2}. Spin connection is related to Affine connection $\Gamma^{i}_{jk}$
and torsion $T^a_{.\mu \nu}$ as follows:
\begin{eqnarray}
 \Gamma^{a}_{\mu
\nu} &=&  C^{a}_{ . \mu\nu}, \quad T^a_{.\mu \nu}  =  C^{a}_{ .[\mu\nu]}
\end{eqnarray}
Now let us introduce the irreducible components of torsion \cite{Shapiro}:
\begin{eqnarray}
S^i& =& \epsilon^{jkli}T_{jkl}, \quad
T_i = T^j_{.ij}, \quad
T_{ijk} = \frac{1}{3}(T_j \eta_{ik}-T_k\eta_{ij}) -
\frac{1}{6}\epsilon_{ijkl}S^l + q_{ijk}
\end{eqnarray}
$S_T$ may be represented in terms of torsion:
\begin{eqnarray}
S_T =  M_T^2 \int \{\frac{2}{3}T^2 - \frac{1}{24}S^2 \}d^4x + \tilde{S}
\end{eqnarray}
Here $\tilde{S}$ depends on $q$ but does not depend on $S, T$.

\section{Massless fermions}

We consider the action of a massless Dirac spinor coupled to the $SO(3,1)$ gauge field in
the form \cite{Alexandrov}:

\begin{eqnarray}
S_f & = & \frac{i}{2}\int  \{ \bar{\psi} \gamma^{\mu} \zeta
D_{\mu} \psi - [D_{\mu}\bar{\psi}]\bar{\zeta}
\gamma^{\mu}\psi \} d^4 x \label{Sf}
\end{eqnarray}
Here $\zeta=\eta + i\theta$ is the coupling constant. 
Let us restore the field of the vierbein that corresponds to vanishing Riemannian curvature. The symmetry group of this system is $Diff \otimes SO(3,1)_{local}$. Fixing the trivial value of the vierbein $e^a_k = \delta^a_k$ we break this symmetry: $Diff \otimes SO(3,1)_{local} \rightarrow SO(3,1)_{local}$. This breakdown is accompanied by distinguishing local Lorentz transformations out of the general coordinate transformations.

In terms of $S$ and $T$ (\ref{Sf}) can be rewritten as:
\begin{eqnarray}
S_f & = & \frac{1}{2}\int \{i\bar{\psi} \gamma^{\mu}\eta
\nabla_{\mu} \psi - i[\nabla_{\mu}\bar{\psi}]{\eta} \gamma^{\mu}\psi  +
\frac{1}{4}\bar{\psi}[\gamma^5 \gamma_d \eta S^d  - 4\theta T^b \gamma_{b}] \psi\} d^4 x \label{Sf22}
\end{eqnarray}

One can see that $\theta$ disappears from the first term with usual derivatives. However, it remains in the second term.
Let us suppose for a moment that it is possible to neglect $S_G$ in some approximation. Then the  integration over torsion degrees of freedom can be performed for the system that consists of the Dirac fermion
coupled to the gauge field. The result of this integration is \cite{Alexandrov}:
\begin{eqnarray}
S_{eff}& = & \frac{1}{2}\int \{i\bar{\psi} \gamma^{\mu} \eta
\nabla_{\mu} \psi - i[\nabla_{\mu}\bar{\psi}] \eta \gamma^{\mu}\psi\}d^4x -
\frac{3}{32M_T^2} \int
\{V^2 \theta^2 - A^2\eta^2 \} d^4x \label{F42}
\end{eqnarray}
Here we have defined $V_{\mu}  =  \bar{\psi} \gamma_{\mu}  \psi$, $A_{\mu}  = \bar{\psi}\gamma^5 \gamma_{\mu}  \psi$. 
The four fermion effective action appears (see also \cite{Xue,Alexander1,Alexander2}).

The next step is the consideration of the fermions coupled to the gauge field of Lorentz group
 with action $S_T+S_G$. This theory may appear to be  renormalizable due to the presence of terms quadratic
 in curvature \cite{Sesgin,Elizalde}.  The effective charges entering the term of the action with the
 derivative of torsion depend on the ratio $\epsilon/\Lambda_{\chi}$, where $\epsilon$ is the energy scale of the considered physical process.
We suppose that under certain circumstances running coupling constants $\beta_i(\epsilon/\Lambda_{\chi})$ are decreased with the decrease of $\epsilon$. Due to the existence of $S_T$ in the action we do not expect confinement, and at small enough energies $\epsilon << \Lambda_{\chi}$ it is possible to neglect $S_G$.

\section{Chiral symmetry breaking }

We restrict ourselves with the one - loop analysis of the effective NJL model. This is based on the assumption that due to a certain symmetry the dominant divergencies that appear at more than one loop are exactly cancelled by contribution of the trans - $\Lambda_{\chi}$ degrees of freedom of the microscopic theory \cite{quantum_hydro,hydro_gravity}. As a result of the cancellation  we are able to use the one - loop gap equation for the consideration of the chiral symmetry breaking.

 Let us
 apply Fierz transformation to the four fermion term of (\ref{F42}):
\begin{eqnarray}
S_{4}  &=&  \frac{3}{32M_T^2}\int \{ -\theta^2(\bar{\psi}^a\gamma^i
\psi^a)(\bar{\psi}^b\gamma_i \psi^b)\} d^4 x
+ \eta^2(\bar{\psi}^a\gamma^i \gamma^5
\psi^a)(\bar{\psi}^b\gamma_i \gamma^5  \psi^b)\} d^4 x\nonumber\\
&=& \frac{3}{32M_T^2}\int\{4 (\eta^2+\theta^2)
(\bar{\psi}^a_{L}\psi^b_{R})(\bar{\psi}^b_{R} \psi^a_{L})
+(\eta^2-\theta^2)[(\bar{\psi}^a_{L}\gamma_i\psi^b_{L})(\bar{\psi}^b_{L}\gamma^i
\psi^a_{t,L})\nonumber\\&&+(L\leftarrow \rightarrow R)]\} d^4 x \label{S4}
\end{eqnarray}
In this form the action is equal to that of the extended NJL model (see Eq. (4), Eq. (5), Eq. (6) of \cite{ENJL}). In
the absence of the other gauge fields the $SU({ N})_L\otimes
SU({ N})_R$ symmetry is broken down to $SU({N})_V$
(here ${ N}$ is the total number of fermions). We suppose that the Electroweak interactions act
as a perturbation.
 We
denote $G_S = \frac{3 (\theta^2+\eta^2)
 \Lambda^2_{}}{64 M_T^2\pi^2}$. Here
$\Lambda_{}$ is the cutoff that is now the physical parameter of the model.
 The gap equation
is
\begin{equation}
m  \approx G_S m (1 - \frac{m^2}{\Lambda_{}^2}{\rm log} \frac{\Lambda_{}^2}{m^2}), \quad (\Lambda_{} \gg m)  \label{gap}
\end{equation}
There exists the critical value of
$G_S$: at $G_S > 1$ the gap equation has the nonzero solution for $m$ while
for $G_S < 1$ it has not.
The analogue of the technipion decay constant $F_T$ in the given approximation is $F^2_T \approx \frac{m^2}{8\pi^2} \, {\rm log}\, \frac{\Lambda_{}^2}{m^2}$.
In order to have appropriate values of $W$ and $Z$ - boson masses we need
$F_T\sim 250/\sqrt{N}$ Gev, where $N=24$ is the total number of fermions.  Rough estimate with $m = 174$ GeV (top quark mass)  gives  the value $\Lambda_{}\sim 5$ TeV. At the same time if only one $t$ quark contributes to the formation of the gauge boson masses we would have $\Lambda_{}\sim 10^{15}$ GeV.
  In order to avoid the constraints on the FCNC we need $\Lambda_{} \ge 1000$ TeV.

\section{Lattice discretization}

 It is implied that the Wick rotation is performed. Therefore, we deal with the Euclidean theory and with the gauge group $SO(4)$ instead of $SO(3,1)$.  The $SO(4)$ connection is attached to links. The curvature is localized on the plaquettes.
The model to be discretized has the action $S = S_f + S_T + S_G$, where the different terms are
given by Eq. (\ref{Sf}), Eq. (\ref{ST}), Eq. (\ref{ST2}).
The lattice discretized action for the fermions is similar to that of suggested in \cite{Diakonov} (we performed the rotation to the Euclidean signature of space - time):
\begin{eqnarray}
S_f & = & \sum_{xy} \{ {\psi}^+_{L,x} H^{xy}_L \psi_{L,y} + \psi^+_{R,x} H^{xy}_R \psi_{R,y} \}, \label{Sfl}
\end{eqnarray}
where
\begin{eqnarray}
H_L & = & \sum_k\Bigl( \eta\{ U^+_{R,yx} \bar{\tau}^k \delta_{x-e_k,y} - \bar{\tau}^k U_{L,xy}  \delta_{x+e_k,y} \}-i\theta\{ U^+_{R,yx} \bar{\tau}^k \delta_{x-e_k,y} + \bar{\tau}^k  U_{L,xy} \delta_{x+e_k,y} - 2 \bar{\tau}^k \delta_{x,y} \}\Bigr)\nonumber\\
H_R & = & \sum_k \Bigl(\eta\{ U^+_{L,yx} {\tau}^k \delta_{x-e_k,y} - {\tau}^kU_{R,xy}  \delta_{x+e_k,y} \} -i\theta\{ U^+_{L,yx} {\tau}^k \delta_{x-e_k,y} + {\tau}^k U_{R,xy}  \delta_{x+e_k,y} - 2 {\tau}^k \delta_{x,y} \}\Bigr)\nonumber\\
&& \tau^4 = \bar{\tau}^4 = -i, \quad \tau^a = -\bar{\tau}^a = \sigma^a (a=1,2,3) \label{Sfl_}
\end{eqnarray}
Here the link matrix is $U_{yx} = \left(\begin{array}{cc} U_{L,yx} & 0 \\
 0 & U_{R,yx} \end{array}\right), \quad U_L,U_R \in SU(2)$.
The gauge symmetry is broken by the lattice discretization, and is restored in the continuum limit, when the invariance under local $SO(4)$ coordinate  transformations comes back.
The fermion action  can also be rewritten in the following way:
\begin{eqnarray}
S_f & = & \sum_{xy}  {\psi}^+_{x} \Gamma^4 {\cal H}^{xy}_L \psi_{y}, \quad
{\cal H} = \left(\begin{array}{cc} 0 & H_R \\
 H_L & 0 \end{array}\right), \quad H_L^+ = H_R \label{HH}
\end{eqnarray}

For $\theta\ne 0$ in the absence of the gauge field the given discretization gives no doublers just like the Wilson formulation of lattice fermions.

Next, let us introduce the definition of lattice $SO(4)$ curvature. It corresponds to the closed path along the boundary of the given plaquette  started from the given lattice site $x$.  Therefore, it is marked by the position of the lattice site $x$ and the couple of the directions in space - time $n,j$.
\begin{eqnarray}
{\cal G}_{x,n,j}^{4,k}  & = & - {\cal G}_{x,n,j}^{k,4} = i \,{\rm sign}(n)\, {\rm sign}(j) \, \Bigl(  {\rm Tr}\, ( U_{L,x,n,j} - U^+_{L,x,n,j}) \sigma^k  -  {\rm Tr}\, ( U_{R,x,n,j} - U^+_{R,x,n,j}) \sigma^k \Bigr) \nonumber\\
{\cal G}_{x,n,j}^{k,l} & = &   i\, {\rm sign}(n)\, {\rm sign}(j) \, \epsilon^{klm} \Bigl( {\rm Tr}\, ( U_{L,x,n,j} - U^+_{L,x,n,j}) \sigma^m +  {\rm Tr}\, ( U_{R,x,n,j} - U^+_{R,x,n,j}) \sigma^m \Bigr)
\end{eqnarray}
Here  $n,j = \pm 1,\pm 2,\pm 3,\pm 4$. Positive sign corresponds to the positive direction while negative sign corresponds to the negative direction in $4D$ space - time. The plaquette variables obtained by the product of link matrices along the boundary of the plaquette located in $n,j$ plane (starting from the point $x$) are denoted by $U_{R,x,n,j}$ and $U_{R,x,n,j}$. Contraction of indices results in the definition of lattice Ricci tensor $\cal R$ and the lattice scalar curvature $\cal S$. Both these quantities are also attached to the closed paths around the boundaries of the plaquettes and are marked by $x, n, j$.
\begin{eqnarray}
{\cal R}_{x,n,j}^{k}  & = &  {\cal G}_{x,n,j}^{k,|n|}, \quad {\cal S}_{x,n,j}   =    {\cal R}_{x,n,j}^{|j|}
\end{eqnarray}

For the part of the action linear in curvature we have $S_T  =  -\kappa \sum_{x} \sum_{n,j=\pm 1,\pm 2, \pm 3, \pm 4} {\cal S}_{x, n, j}$.
Analogue of quantity $A$  is given by
${\cal A}_{x,n,j}   =   \sum_{kl}\epsilon^{|n||j|kl} {\cal G}_{x,n,j}^{k,l}$.
  Contraction of indices in quadratic expressions results in the following combinations:
\begin{eqnarray}
Q_{x}^{(1)}  & = &  \sum_{k,l, n, j} {\cal G}_{x,n,j}^{|k|,|l|}{\cal G}_{x,n,j}^{|k|,|l|}, \quad
 Q^{(2)}_x = \sum_{k,l, n, j} {\cal G}_{x,n,j}^{|k|,|l|}{\cal G}_{x,k,l}^{|n|,|j|}, \quad
 Q^{(3)}_x = \sum_{k, n, j} {\cal R}_{x,n,j}^{|k|}{\cal R}_{x,n,j}^{|k|},\nonumber\\
 Q^{(4)}_x &=& \sum_{k, n, j} {\cal R}_{x,n,j}^{|k|}{\cal R}_{x,n,k}^{|j|}, \quad
 Q^{(5)}_x = \sum_{n, j} {\cal S}_{x,n,j}{\cal S}_{x,n,j}, \quad
 Q^{(6)}_x = \sum_{ n, j} {\cal A}_{x,n,j}{\cal A}_{x,n,j}
\end{eqnarray}
(The sum is over positive and negative values of $n,j,k,l$.)
Finally, the term in the action quadratic in curvature has the form \cite{Z2006}
\begin{eqnarray}
S_G & = & \sum_{x}\sum_{i=1}^6 \beta_i Q^{(i)}_x
\end{eqnarray}


Our main expectation is that in this theory the chiral symmetry is broken dynamically while the confinement is absent.
The main quantities to be measured are the static potential extracted from the Polyakov loops and the chiral condensate.
 The chiral condensate $\chi$ may be calculated using the Banks - Casher relation $\chi  =  - \frac{\pi}{V} \langle \nu(0)\rangle$,
where $\nu(\lambda)$ is the density of eigenvalues of the operator $\cal H$ given by Eq. (\ref{HH}), $V$ is the $4D$ volume.
The potential $V_{L\bar{L}}(R)=V_{R\bar{R}}(R)$ between the static (either left - handed or right - handed) fermion and anti - fermion is defined through the relation
\begin{eqnarray}
&&{\rm exp}( - V_{L\bar{L}}(|x-y|) T)  =   \langle {\cal P}_L(x) {\cal P}_R(y) \rangle, \\
&& {\cal P}_L(x) = {\rm Tr} \Bigl(\Pi_{K = 0, ... , T-1}U_{L, x + Ke_4, x+(K+1)e_4}\Bigr) , \quad  {\cal P}_R(y) = {\rm Tr} \Bigl(\Pi_{K = 0, ... , T-1}U_{R, x + Ke_4, x+(K+1)e_4}\Bigr)\nonumber
\end{eqnarray}
Here $x_4 = y_4 = 0$, while $T$ is the lattice extent in time direction.
The potential $V_{L\bar{R}}(R)=V_{R\bar{L}}(R)$ between the static fermion and anti - fermion of different chiralities is given by \begin{equation}
{\rm exp}( - V_{L\bar{R}}(|x-y|) T) = \langle {\cal P}_L(x) {\cal P}_L(y)\rangle = \langle {\cal P}_R(x) {\cal P}_R(y)\rangle
\end{equation}

\section{Conclusions}

 We have suggested the possible new scenario of the dynamical electroweak symmetry breaking. In this scenario Lorentz group plays the role of the technicolor group. We do not introduce the additional technifermions. Instead, the Higgs bosons are composed directly of the existing Standard Model fermions. The chiral symmetry breaking may take place in the given model. This is demonstrated in the approximation when the squared curvature gauge field action $S_G$ is neglected. In the same approximation there is no confinement. This means, that in the complete theory with $S_G$ taken into account the chiral symmetry breaking may take place in the absence of confinement. If so, the chiral condensate provides masses for the electroweak gauge bosons while all fermions have equal masses. However, small perturbations to this pattern may cause the difference between the masses of fermions \cite{Zubkov:2013zg}. We suppose, that the observed  massive Standard Model fermions may appear in this way.

We have suggested the lattice discretization of the Euclidean version of the model with the gauge field from $SO(4)\approx SU(2)\otimes SU(2)$. In lattice formulation the gauge invariance is lost and is supposed to be restored in the continuum limit. In the presence of nonminimal interactions the fermion doublers disappear. The Dirac operator is Hermitian, and the Banks - Casher relation can be used in order to check the appearance of the chiral condensate.

The author is greatful to D.I. Diakonov,  Yu.A. Simonov,  B. Svetitsky, G.E. Volovik, and to V.I. Zakharov for useful discussions.    This work was partly supported by RFBR grant 11-02-01227, by the
Federal Special-Purpose Programme 'Human Capital' of the Russian Ministry of
Science and Education.


\begin{thebibliography}{99}




\bibitem{Shapiro}
A.S.Belyaev, I.L.Shapiro, Nucl.Phys. B543 (1999) 20-46, ArXiv:hep-ph/9806313

\bibitem{Elizalde}
E. Elizalde, S.D.Odintsov, Int.J.Mod.Phys.D2:51-58,1993


\bibitem{Minkowski}
Nakia Carlevaro, Orchidea Maria Lecian, Giovanni Montani, Int. J. Mod. Phys. A
23, 1282-1285 (2008)

\bibitem{Minkowski_}
Nakia Carlevaro, Orchidea Maria Lecian, Giovanni Montani,
Mod.Phys.Lett.A24:415-427,2009

\bibitem{Z2010_3}
M.A.Zubkov, Mod. Phys. Lett. A25:2885-2898,2010, ArXiv:1003.5473


\bibitem{topcolor1}
William A. Bardeen, Christopher T. Hill, Manfred Lindner,
"Minimal Dynamical Symmetry Breaking of the Standard Model,"
Phys. Rev. D {\bf 41}, 1647--1660 (1990).


\bibitem{Miransky}
V.A. Miransky,  Masaharu Tanabashi, Koichi Yamawaki,
``Dynamical electroweak symmetry breaking with large anomalous
dimension and t quark condensate",
Phys. Lett. B {\bf 221}, 177--183 (1989).

``Is the t quark responsible for the mass of W and Z bosons?",
Mod.  Phys. Lett. A {\bf 4}, 1043--1053  (1989).

\bibitem{Simmons}
Christopher T. Hill, Elizabeth H. Simmons,
"Strong Dynamics and Electroweak Symmetry Breaking",
Phys. Rept. {\bf 381}, 235--402 (2003) ; Erratum-ibid. {\bf 390}, 553--554 (2004).


\bibitem{NJL}
Y. Nambu, G. Jona-Lasinio,
"Dynamical model of elementary particles based on an analogy with superconductivity. I,"
Phys. Rev. {\bf 122}, 345--358 (1961).


\bibitem{VZ2013}
G.~E.~Volovik and M.~A.~Zubkov,
 arXiv:1302.2360, Pisma v ZhETF, vol. 97 (2013), issue 6, page 344

\bibitem{VZ2012}
G.~E.~Volovik and M.~A.~Zubkov,
 Phys. Rev. D 87, 075016 (2013),  arXiv:1209.0204

\bibitem{VZ2013_2} G.E.Volovik, M.A.Zubkov, 	arXiv:1305.7219, to appear in Journal of Low temperature Physics

\bibitem{Zubkov:2013zg}
  M.~A.~Zubkov,
  ``Gauge theory of Lorentz group as a source of the dynamical electroweak symmetry breaking,''
  arXiv:1301.6971 [hep-lat], to appear in JHEP














\bibitem{quantum_hydro}
G.E. Volovik, "From Quantum Hydrodynamics to Quantum Gravity", arXiv:gr-qc/0612134,
 Proceedings of the Eleventh Marcel Grossmann Meeting on General Relativity, edited by H. Kleinert, R.T. Jantzen and R. Ruffini, World Scientific, Singapore, 2008, pp. 1404-1423

\bibitem{hydro_gravity}
G.E. Volovik, "Vacuum energy: quantum hydrodynamics vs quantum gravity", arXiv:gr-qc/0505104,  JETP Lett. 82 (2005) 319-324; Pisma Zh.Eksp.Teor.Fiz. 82 (2005) 358-363



\bibitem{Alexandrov}
 Sergei Alexandrov,
Class.Quant.Grav.25:145012,2008

\bibitem{Xue}
She-Sheng Xue, Phys.Lett.B665:54-57,2008, ArXiv:0804.4619

\bibitem{Alexander1}
S.Alexander, T.Biswas, G.Calcagni, Phys. Rev. D 81, 043511 (2010),
ArXiv:0906.5161

\bibitem{Alexander2}
S.Alexander, D.Vaid, ArXiv:hep-th/0609066












\bibitem{Diakonov}
A.~A.~Vladimirov and D.~Diakonov,
{ ``Phase transitions in spinor quantum gravity on a lattice''},
  {}arXiv:1208.1254 [hep-th]
  {}10.1103/PhysRevD.86.104019,
{}Phys.\ Rev.\ D {\bf 86}, 104019 (2012) 

{}D.~Diakonov,
{ ``Towards lattice-regularized Quantum Gravity''},
  {}arXiv:1109.0091 [hep-th]
\bibitem{Z2006}
M.A.Zubkov, { ``Gauge invariant discretization of Poincare quantum gravity''},
 {}Phys.\ Lett.\  B {\bf 638}, 503 (2006),
  [Erratum-ibid.\  B {\bf 655}, 307 (2007)]
  [arXiv:hep-lat/0604011]



\bibitem{ENJL}
J.Bijnens, C.Bruno, E. de Rafael, Nucl.Phys. B390 (1993) 501-541, hep-ph/920623

\bibitem{Diakonov2}
D.~Diakonov, A.~G.~Tumanov and A.~A.~Vladimirov,
{ ``Low-energy General Relativity with torsion: A Systematic derivative expansion''},
{}arXiv:1104.2432 [hep-th], {}10.1103/PhysRevD.84.124042, {}Phys.\ Rev.\ D {\bf 84}, 124042 (2011) 


\bibitem{ConformalGrav}
 Philip D. Mannheim, Prog.Part.Nucl.Phys. 56 (2006) 340-445

V.V. Zhytnikov, Int.J.Mod.Phys.A8:5141-5152,1993.


\bibitem{Sesgin}
E. Sezgin, P. van Nieuwenhuizen, Phys.Rev.D22:301,1980.








\end{thebibliography}
\end{document}